\documentclass[twocolumn]{jpsj3}
\usepackage{txfonts}
\usepackage{color}
\usepackage{ulem}

\usepackage{color}
\newcommand{\Tc}{T_{\rm c}}

\newcommand{\LaTAl}{La$\it Tr\rm_{2}$Al$_{20}$}

\newcommand{\RxVAl}{$\it R_{x}$V$_{2}$Al$_{20}$}

\title{Superconductivity in Cage Compounds La$\it Tr\rm_{2}$Al$_{20}$ with $\it Tr$ = Ti, V, Nb, and Ta}

\author{Akira Yamada, Ryuji Higashinaka, Tatsuma D. Matsuda, and Yuji Aoki~\thanks{aoki@tmu.ac.jp}} 
\inst{Department of Physics, Tokyo Metropolitan University, Hachioji, Tokyo 192-0397, Japan} 

\abst{Electrical resistivity, magnetic susceptibility, and specific heat measurements on single crystals of {\LaTAl} ($\it Tr\rm$ = Ti, V, Nb, and Ta) revealed that these four compounds exhibit weak-coupling superconductivity with transition temperatures $\Tc$ = 0.46, 0.15, 1.05, and 1.03 K, respectively. %
LaTi$_{2}$Al$_{20}$ is most probably a type-I superconductor, which is quite rare among intermetallic compounds. %
Single-crystal X-ray diffraction suggests ``rattling" anharmonic large-amplitude oscillations of Al ions (16$c$ site) on the Al$_{16}$ cage, while no such feature is suggested for the cage-center La ion.
Using a parameter $d_{\rm GFS}$ quantifying the ``guest free space" of the cage-center ion, we demonstrate that nonmagnetic $\it RTr_{2}$Al$_{20}$ superconductors are classified into two groups, i.e., (A) $d_{\rm GFS} \ne 0$ and $\Tc$ correlates with $d_{\rm GFS}$, and (B) $d_{\rm GFS} \simeq 0$ and $\Tc$ seems to be governed by other factors.
}

\begin{document}
\maketitle
\newpage
Intermetallic compounds $\it RTr\rm_{2}\it X\rm_{20}$ ($\it R\rm$ : rare earth, $\it Tr$ : transition metals, $\it X\rm$ : Al, Zn, and Cd), which crystallize in the cubic CeCr$_{2}$Al$_{20}$-type structure ($Fd\bar{3}m$, $\#227$), have attracted considerable attention in recent years, because a wide variety of strongly correlated electron phenomena caused by strong $c$-$f$ hybridization have been observed.
YbCo$_{2}$Zn$_{20}$ is a heavy fermion (HF) compound with an electronic specific heat coefficient of 8 J/(mol K$^{2}$), which is the largest among Yb compounds~\cite{Torikachvili_PNAS_07}.
Sm$\it Tr\rm_{2}$Al$_{20}$ ($\it Tr$ = Ti, V, Cr, and Ta) exhibit rare Sm-based HF behaviors, which are anomalously field-insensitive~\cite{Higashinaka_JPSJ_11_SmTi2Al20, Sakai_PRB_11, Yamada_JPSJ_13}.
Many of the Pr$\it Tr\rm_{2}\it X\rm_{20}$ compounds have a non-Kramers $\Gamma_{3}$ doublet crystalline-electric-field ground state, and exhibit quadrupole Kondo lattice behaviors~\cite{Sakai_JPSJ_11, Onimaru_JPSJ_16, Yoshida_JPSJ_17, Higashinaka_JPSJ_17}.
Therefore, it is presumed that the superconductivity (SC) appearing in the Pr$\it Tr\rm_{2}\it X\rm_{20}$ compounds is induced by quadrupolar fluctuations~\cite{Onimaru_JPSJ_11, Sakai_JPSJ_12_PrTi2Al20, Matsubayashi_PRL_12, Tsujimoto_PRL_14}.

The SC appearing in $\it RTr\rm_{2}\it X\rm_{20}$ with nonmagnetic $R$ ions has been discussed in terms of the cage structure, which is one of the characteristic features of the CeCr$_{2}$Al$_{20}$-type crystal structure.
The $\it R$ ions at the $8a$ site with cubic $\it T\rm_{d}$ symmetry are located at the center of an $\it X\rm_{16}$ cage (CN 16 Frank-Kasper polyhedron).
In {\RxVAl} with $\it R$ = Al and Ga (the SC transition temperatures $\Tc$ are 1.49 and 1.66 K, respectively), the cage-center $R$ ions show anharmonic large-amplitude oscillations, which are considered to enhance $\Tc$ through the electron-phonon coupling~\cite{Hiroi_JPSJ_12, Onosaka_JPSJ_12, Safarik_PRB_12, Koza_PCCP_14}.
Similar discussions have also been made for $\it R$ = Sc, Y, Lu, and La~\cite{Winiarski_PRB_16}.

In contrast, compounds with $\it X\rm$ = Zn have different characteristics of lattice oscillations.
La$\it Tr\rm_{2}$Zn$_{20}$ with $\it Tr$ = Ru, Ir, and Os show structural phase transitions at 150, 200, and 151 K, followed by SC transitions at 0.2, 0.6, and 0.07 K, respectively~\cite{Onimaru_JPSJ_10, Wakiya_JPSJ_17}.
Inelastic X-ray scattering measurements and a first-principle calculation suggest that the structural transitions are associated with the low-frequency vibrations of Zn at the $16c$ site on the cage~\cite{Hasegawa_JPCS_12, Wakiya_PRB_16}.
The relation between the structural phase transition and SC in these compounds is yet to be understood.

For $\it X\rm$ = Al, the physical properties have not been fully investigated yet, except $\it Tr =$ V mentioned above.
In this paper, we report our studies on {\LaTAl} ($\it Tr\rm$ = Ti, V, Nb, and Ta) using single crystals, which reveal that these compounds exhibit superconductivity.


Single crystals of La$\it Tr\rm_{2}$Al$_{20}$ ($\it Tr\rm =$ Ti, V, Nb, and Ta) were grown using the self-Al flux method. The starting materials were La chips (99.9$\%$), Al grains (99.99$\%$) and powders of Ti (99.99$\%$), V (99.9$\%$), Nb (99.9$\%$) and Ta (99.95$\%$). With an atomic ratio of La:$\it Tr\rm$:Al = 1:2:90, the starting materials were put in an alumina crucible and sealed in a quartz tube.
The quartz tube was heated to 1050 $^{\rm o}$C and then was slowly cooled.
Single crystals were obtained by spinning the ampoule in a centrifuge in order to remove the excess Al flux. 

The electrical resistance and specific heat were measured using a Quantum Design (QD) Physical Property Measurement System (PPMS) equipped with a Helium-3 cryostat and an adiabatic demagnetization refrigerator (ADR).
The ac magnetic susceptibility $\chi_{\rm ac}$ was measured by a mutual inductance method~\cite{chi_ac} down to 0.18 K with a modulation field of $H_{\rm ac}$ = 0.3 Oe using a $^{3}$He-$^{4}$He dilution refrigerator.

\begin{table}[tb]
\caption{Crystallographic parameters of La$\it Tr\rm_2$Al$_{20}$ ($\it Tr\rm =$ Ti, V, Nb, and Ta) at room temperature. $R$ and $wR$ are reliability factors. $B_{\rm eq}$ is the equivalent isotropic atomic displacement parameter. Standard deviations in the positions of the least significant digits are given in parentheses.}
\label{tablestruct}
\begin{center}
\begin{tabular}{lccccc}
\hline
\multicolumn{3}{c}{LaTi$_{2}$Al$_{20}$} & \multicolumn{3}{l}{$R$ $=$ 2.46$\%$, $wR$ $=$ 4.88$\%$}\\
\multicolumn{3}{c}{$Fd\bar3m$ ($\sharp$227) (origin choice 2)} &\multicolumn{3}{l}{$a$ $=$ 14.7946(15) $\AA$, $V$ $= $ 3238.2(6) $\AA^3$}\\
\multicolumn{2}{l}{} & \multicolumn{3}{c}{Position}\\
\cline{3-5}
Atom & site & $x$ & $y$ & $z$ & $B_{\rm eq}$($\AA^2$)\\
\hline
La & $8a$ & 1/8 & 1/8 & 1/8 & 0.62(2) \\
Ti & $16d$ & 1/2 & 1/2 & 1/2 & 0.39(3) \\
Al(1) & $96g$ & 0.05910(6) & 0.05910(6) & 0.3257(8) & 0.88(2) \\
Al(2) & $48f$ & 0.48709(12) & 1/8 & 1/8 & 0.67(3) \\
Al(3) & $16c$ & 0 & 0 & 0 & 1.65(6) \\
\hline
\\
\hline
\multicolumn{3}{c}{LaV$_{2}$Al$_{20}$} & \multicolumn{3}{c}{$R$ $=$ 2.30$\%$, $wR$ $=$ 4.96$\%$}\\
\multicolumn{3}{c}{$Fd\bar3m$ ($\sharp$227) (origin choice 2)} & \multicolumn{3}{c}{$a$ $=$ 14.6125(15) $\AA$, $V$ $= $ 3120.1(6) $\AA^3$}\\
\multicolumn{2}{l}{} & \multicolumn{3}{c}{Position}\\
\cline{3-5}
Atom & site & $x$ & $y$ & $z$ & $B_{\rm eq}$($\AA^2$)\\
\hline
La & $8a$ & 1/8 & 1/8 & 1/8 & 0.72(2) \\
V & $16d$ & 1/2 & 1/2 & 1/2 & 0.29(3) \\
Al(1) & $96g$ & 0.05871(8) & 0.05871(8) & 0.32618(6) & 0.71(2) \\
Al(2) & $48f$ & 0.4871(11) & 1/8 & 1/8 & 0.52(3) \\
Al(3) & $16c$ & 0 & 0 & 0 & 1.57(4) \\
\hline
\\
\hline
\multicolumn{3}{c}{LaNb$_{2}$Al$_{20}$} & \multicolumn{3}{c}{$R$ $=$ 2.44$\%$, $wR$ $=$ 4.94$\%$}\\
\multicolumn{3}{c}{$Fd\bar3m$ ($\sharp$227) (origin choice 2)} & \multicolumn{3}{c}{$a$ $=$ 14.8180(15) $\AA$, $V$ $= $ 3253.6(6) $\AA^3$}\\
\multicolumn{2}{l}{} & \multicolumn{3}{c}{Position}\\
\cline{3-5}
Atom & site & $x$ & $y$ & $z$ & $B_{\rm eq}$($\AA^2$)\\
\hline
La & $8a$ & 1/8 & 1/8 & 1/8 & 0.66(2) \\
Nb & $16d$ & 1/2 & 1/2 & 1/2 & 0.41(2) \\
Al(1) & $96g$ & 0.05949(7) & 0.05949(7) & 0.32431(10) & 0.88(3) \\
Al(2) & $48f$ & 0.48547(14) & 1/8 & 1/8 & 0.74(4) \\
Al(3) & $16c$ & 0 & 0 & 0 & 1.77(7) \\
\hline
\\
\hline
\multicolumn{3}{c}{LaTa$_{2}$Al$_{20}$} & \multicolumn{3}{c}{$R$ $=$ 1.62$\%$, $wR$ $=$ 3.45$\%$}\\
\multicolumn{3}{c}{$Fd\bar3m$ ($\sharp$227) (origin choice 2)} & \multicolumn{3}{c}{$a$ $=$ 14.8231(13) $\AA$, $V$ $= $ 3257.0(5) $\AA^3$}\\
\multicolumn{2}{c}{} & \multicolumn{3}{c}{Position}\\
\cline{3-5}
Atom & site & $x$ & $y$ & $z$ & $B_{\rm eq}$($\AA^2$)\\
\hline
La & $8a$ & 1/8 & 1/8 & 1/8 & 0.61(2) \\
Ta & $16d$ & 1/2 & 1/2 & 1/2 & 0.332(14) \\
Al(1) & $96g$ & 0.05946(8) & 0.05946(8) & 0.32445(11) & 0.82(3) \\
Al(2) & $48f$ & 0.48536(15) & 1/8 & 1/8 & 0.68(4) \\
Al(3) & $16c$ & 0 & 0 & 0 & 1.57(8) \\
\hline
\end{tabular}
\end{center}
\end{table}

Single crystal X-ray diffraction analysis was performed using a Rigaku XtaLABmini with graphite monochromated Mo-K$_{\rm\alpha}$ radiation.
The structural parameters refined using the program SHELXL~\cite{Sheldrick_SHELX-97_97} are shown in Table~\ref{tablestruct}.
The lattice parameters $a$ are close to those in the previous reports~\cite{Niemann_JSSC_95, Thiede_JAAC_98}.
Note that the equivalent isotropic atomic displacement parameters $B_{\rm eq}$ of Al(3) at the 16$c$ site have relatively large values: $B_{\rm eq}=1.57-1.77$ $\AA^2$.
This feature is characteristic to $\it RTr\rm_{2}\it X\rm_{20}$ compounds; see Refs.\cite{Nasch_ZNB_97, Kangas_JSSC_12} for $\it X\rm$ = Al and Refs.~\cite{Onimaru_JPSJ_11, Hasegawa_JPCS_12, Wakiya_PRB_16} for $\it X\rm$ = Zn.
The cage-center La ions at the 8$a$ site have normal values of $B_{\rm eq}=0.61-0.72$ $\AA^2$, in contrast to {\RxVAl} ($\it R$ = Al and Ga), in which the cage-center $R$ ions are suggested to have anharmonic rattling modes~\cite{Hiroi_JPSJ_12, Onosaka_JPSJ_12, Safarik_PRB_12, Koza_PCCP_14}.

\begin{figure}
\begin{center}
\includegraphics[width=0.9\linewidth]{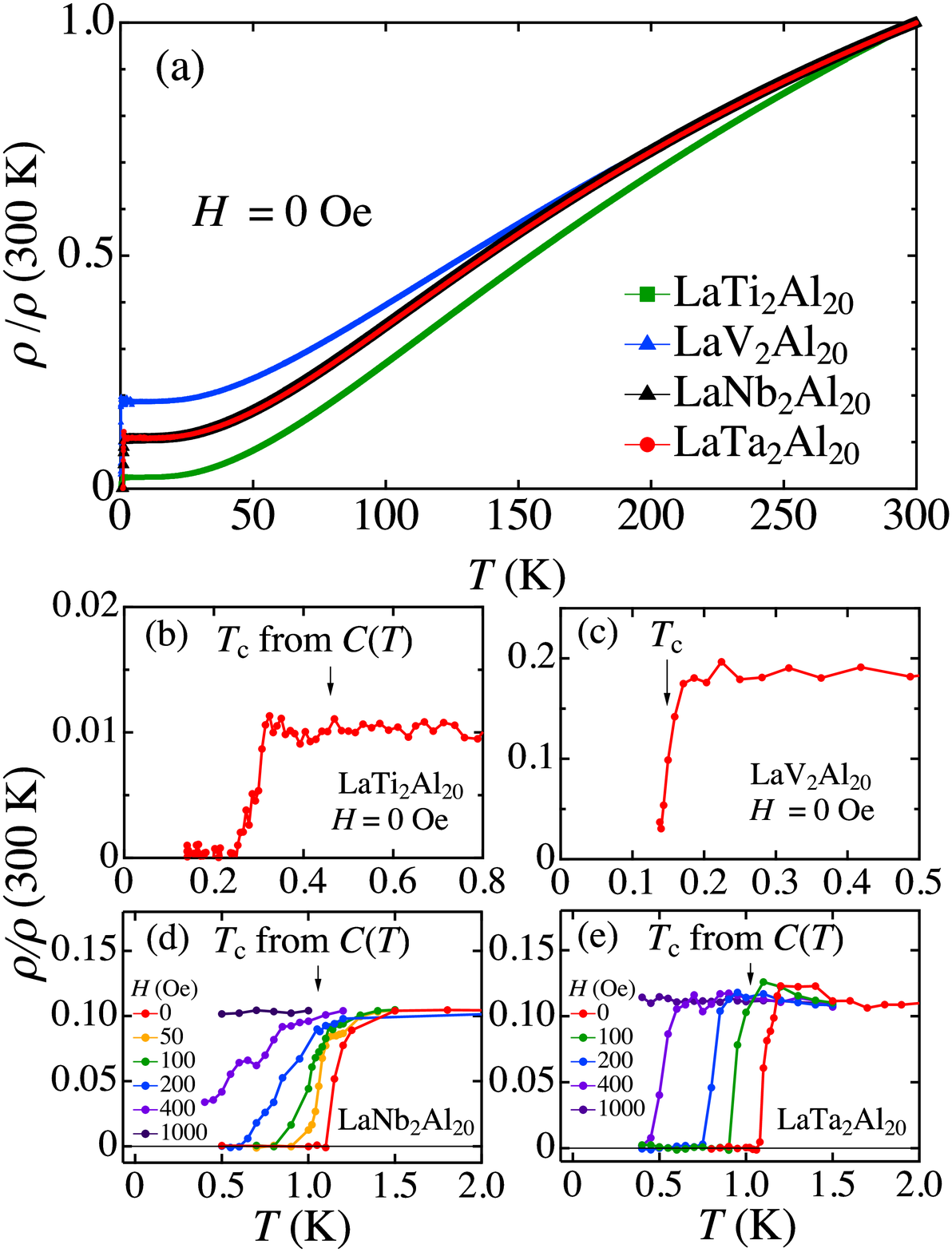}
\end{center}
\caption{(Color online) (a) Temperature dependence of electrical resistivity $\rho$ for {\LaTAl}  ($\it Tr\rm$ = Ti, V, Nb, and Ta) with the current along the [1$\bar{1}$0] direction. (b)-(e) Enlarged views of $\rho(T)$ at low temperatures for each material. Magnetic field is applied along the [111] direction.
For LaTi$_{2}$Al$_{20}$, a drop in $\rho$ appearing at 1.05 K (not shown) is attributable to the SC transition of Al-flux inclusion in the crystal.}
\label{figrho}
\end{figure}

The temperature dependence of resistivity $\rho(T)$ divided by $\rho$(300 K) is shown in Fig.~\ref{figrho}(a).
Note that it is difficult to obtain the absolute values of $\rho(T)$ due to the smallness (typically $0.3 \sim 0.4$ mm in length) of the grown single crystals. %
The residual resistivity ratio $\it RRR\rm$ is $\rho$(300 K)/$\rho$(1.2 K) = 41.2 for Ti, $\rho$(300 K)/$\rho$(1.8 K) = 9.5 for Nb, $\rho$(300 K)/$\rho$(1.8 K) = 9.2 for Ta, and $\rho$(300 K)/$\rho$(0.2 K) = 5.4 for V.
Figures~\ref{figrho}(b-e) show the low-temperature expansion of $\rho(T)$ data for each compound.
In zero field, all the compounds show SC transitions at 0.30, 0.15, 1.15, and 1.10 K (determined by the 50\% transition in $\rho$) for $\it Tr\rm$ = Ti, V, Nb, and Ta, respectively.
These values are slightly different from those determined by the thermodynamic quantities as shown below.
In the applied fields, the transition temperature shifts to lower temperatures.
The superconductivity is completely suppressed in 1 kOe above 0.4 K.

\begin{figure}
\begin{center}
\includegraphics[width=0.8\linewidth]{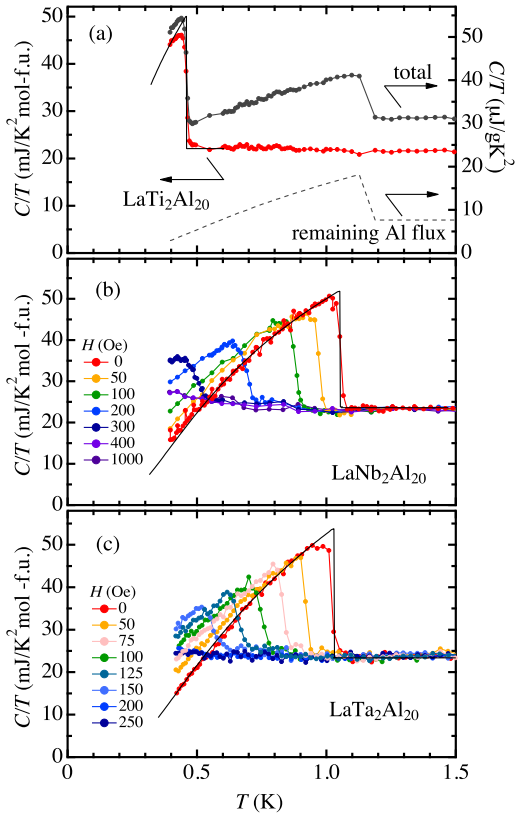}
\end{center}
\caption{(Color online) Temperature dependence of specific heat $C$ divided by temperature for La$\it Tr\rm_{2}$Al$_{20}$ ($\it Tr\rm$ = Ti, Nb, and Ta).
The solid curves represent the fitting by the $\alpha$ model~\cite{Johnson_SST_13, Padamsee_JLTP_73}.
(a) For LaTi$_2$Al$_{20}$, SC transition of the Al-flux remaining inside the crystal is observed at 1.14 K.
The total heat capacity can be expressed as $C_{\rm total} \equiv m_{\rm LaTi_{2}Al_{20}} C_{\rm LaTi_{2}Al_{20}} + m_{\rm Al} C_{\rm Al }$, where $m_{\rm LaTi_{2}Al_{20}} (m_{\rm Al})$ and $C_{\rm LaTi_{2}Al_{20}} (C_{\rm Al})$ are the mass and heat capacity per unit mass for LaTi$_{2}$Al$_{20}$ (Al).
To obtain $C_{\rm LaTi_{2}Al_{20}}$, $m_{\rm Al}$ is determined so that the SC jump of Al inclusion in $C_{\rm total}$ is reproduced by $m_{\rm Al} C_{\rm Al}$.

}
\label{figC}
\end{figure}

Figure~\ref{figC} shows the temperature dependence of specific heat $C$ divided by temperature.
A clear specific heat jump appears at 0.46 K (Ti), 1.05 K (Nb) and 1.03 K (Ta), which is referred to as the bulk SC transition temperature $\Tc$ hereinafter.
Considering that we used 3-7 pieces of single crystals for the measurements, the small transition width indicates high degree of uniformity of the SC phase in the crystals.
For LaTi$_{2}$Al$_{20}$, a jump appears at 1.14 K, indicating Al-flux inclusion in the crystal.
Using $C(T)$ data obtained for Al metal, the contribution from LaTi$_{2}$Al$_{20}$ is separated as shown in Fig.~\ref{figC}.

The normal-state $C/T$ data below 5 K can be described by $C/T = \gamma + \beta T^{2}$, where $\gamma$ and $\beta$ are the electronic and phonon specific heat coefficients, respectively.
The specific heat jump at the SC transition $\Delta C(\Tc$)/$\gamma \Tc$ is 1.26, 1.25, and 1.36 for Ti, Nb, and Ta, respectively.
These values are close to and slightly smaller than 1.43 expected from the BCS theory, indicating that they are weak-coupling superconductors.
The fitting of the $C(T)$ data by the $\alpha$ model~\cite{Johnson_SST_13, Padamsee_JLTP_73} is shown by the thin curve in Fig.~\ref{figC}.
The obtained $\alpha$ value is 1.67, 1.62, and 1.70 for Ti, Nb, and Ta, respectively, which is smaller than 1.764 expected from the BCS theory.
This feature indicates that the size of the SC gap differs slightly in each of the multiple conduction bands.

\begin{figure}
\begin{center}
\includegraphics[width=0.9\linewidth]{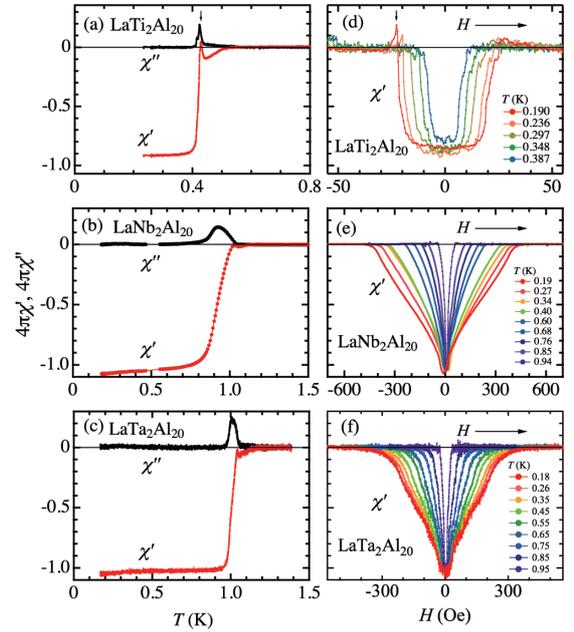}
\end{center}
\caption{(Color online) Temperature dependence (a-c) and magnetic field dependence (d-f) of ac magnetic susceptibility $\chi_{\rm ac} \equiv \chi'-i\chi''$ for La$\it Tr\rm$$_{2}$Al$_{20}$ ($\it Tr\rm$ = Ti, Nb, and Ta). 
The horizontal arrows indicate the direction of the field sweep.
Note that $\chi'$ and $\chi''$ include some errors due to no correction for the demagnetization factor of the crystals.
For LaTi$_2$Al$_{20}$, arrows indicate positive peaks appearing in $\chi'$, ascribable to the ``differential paramagnetic effect (DPE)" (see text).
}
\label{figchiac}
\end{figure}

The temperature dependence of ac magnetic susceptibility $\chi_{\rm ac} \equiv \chi'-i\chi''$ is shown in Figs.~\ref{figchiac} (a-c).
The real-part $\chi'$ shows that the superconducting volume fraction reaches approximately 100$\%$, confirming that the SC in La$\it Tr\rm$$_{2}$Al$_{20}$ is of bulk nature.
Considering that we used 10-20 pieces of single crystals for the measurements, the small transition width indicates high degree of uniformity of the SC phase in the crystals.
Figures~\ref{figchiac} (d-f) show the magnetic field dependence of $\chi'$ at several selected temperatures.
With decreasing temperature, the onset of the SC transition shifts to higher fields.

\begin{table*}[tb] %
\caption{Characteristic parameters of nonmagnetic $\it RTr\rm_{2}$Al$_{20}$ superconductors (see text for definitions).
The errors in the last significant digit(s) are indicated in parentheses.%
The error in $\Tc$ ($\pm \Delta \Tc$) is estimated by the 90-10\% transition in $C/T$.
The standard deviation $\Delta \alpha$ is obtained by least squares fitting of the $\alpha$ model to the $C/T$ data in $T/\Tc<0.9-0.95$ (excluding the SC transition region).
}%
\label{tableSCpara}
\begin{center}
\begin{tabular}{rccccccc}
\hline
compounds & LaTi$_{2}$Al$_{20}$ & LaV$_{2}$Al$_{20}$ & LaNb$_{2}$Al$_{20}$ & LaTa$_{2}$Al$_{20}$ & ScV$_{2}$Al$_{20}$~\cite{Winiarski_PRB_16} & YV$_{2}$Al$_{20}$~\cite{Winiarski_PRB_16} & LuV$_{2}$Al$_{20}$~\cite{Winiarski_PRB_16}\\
\hline
$\Tc$ (K) & 0.46(1) & 0.15(2) & 1.05(2) & 1.03(2) & 1.00 & 0.60 & 0.57\\
$\gamma$ (mJ/molK$^{2}$) & 22 & 19.6~\cite{Winiarski_PRB_16} & 23.4 & 22.1 & 29.68 & 26.46 & 30.05\\
$\gamma_{\rm cal}$ (mJ/molK$^{2}$) & 17.4~\cite{Tanaka_JPSJ_11} & -- & -- & -- & 21.2 & 18.5\\
$\alpha$ & 1.67(1) & -- & 1.62(2) & 1.70(2) & 1.78 & 1.65 & 1.68\\
$\Delta C/\gamma\Tc$ & 1.26  & --  & 1.25 & 1.36 & 1.46 & 1.24 & 1.29\\
$\it\Theta_{\rm D}$ (K) & 510~\cite{Sakai_JPSJ_12} & 525~\cite{Winiarski_PRB_16} & 505 & 480 & 536 & 516 & 502\\
$\lambda_{\rm e\textendash ph}$ & 0.376 & 0.34 & 0.414  & 0.418 & 0.41 & 0.39 & 0.39\\
$H_{\rm c}$(0) (Oe)  & 25 & -- & 57 & 56 & 66 & 36 & 37 \\
$\frac{dH_{\rm c2}}{dT}\vert_{T=T_{\rm c}}$ (Oe/K) & -- & -- & -586 & -390 & -4820 & -670 & -5010 \\ 
$H_{\rm c2}$(0) (Oe)  & -- & -- & 449 & 292 & 3330 & 280 & 2070 \\ 
$\xi_{\rm GL}$ ($\AA$) & -- & -- & 856 & 1060 & 314 & 1084 & 399 \\ 
$\kappa_{\rm GL}=\kappa_2(T \to T_{\rm c})$ & -- & -- & 3.2 & 2.2 & 20.6 & 3.4 & 22.6 \\ 
$\lambda_{\rm L}=\kappa_{\rm GL} \xi_{\rm GL}$ ($\AA$) & -- & -- & 2760 & 2310 & 6450 & 3650 & 9020 \\  
$H_{\rm c1}$(0) (Oe)  & -- & -- & 15 & 14 & 7 & 9 & 4 \\ 
\hline
\end{tabular}
\end{center}
\end{table*}

\begin{figure}
\begin{center}
\includegraphics[width=0.9\linewidth]{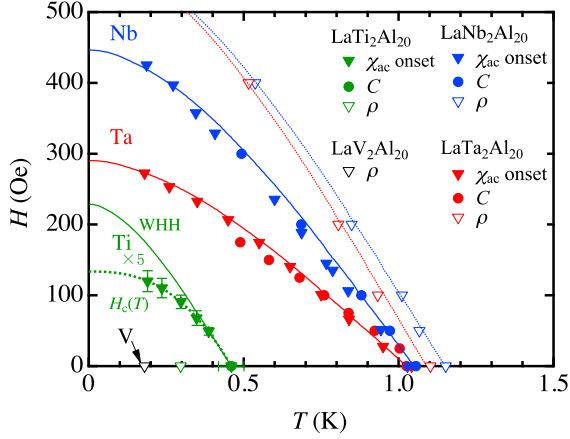}
\end{center}
\caption{(Color online) $H$-$T$ phase diagram of La$\it Tr\rm$$_{2}$Al$_{20}$ (for $\it Tr\rm$ = Ti, $H$ values are scaled by a factor of 5 for clarity).
Solid curves represent the fitting by the Werthamer-Helfand-Hohenberg (WHH) clean-limit model~\cite{HW_PR_66, WHH_PR_66}.
For $\it Tr\rm$ = Ti, the fitting with $H_{\rm c}(T) = H_{\rm c0}[1-(T/T_{\rm c})^s]$ (thick dotted line) reproduces the experimental data better than the WHH model.
For $\it Tr\rm$ = Nb and Ta, the boundary (thin dotted line) determined by the midpoint of the resistive transition is largely enhanced compared to the bulk SC boundary.
This enhancement of the upper critical field $H_{\rm c2}$ might be due to a surface SC effect; theoretically, a surface SC layer can have a upper critical field $H_{\rm c3} = 1.69H_{\rm c2}$~\cite{SJ_dG_PL_64}.
}
\label{figHTdiagram}
\end{figure}

\begin{figure}
\begin{center}
\includegraphics[width=0.95\linewidth]{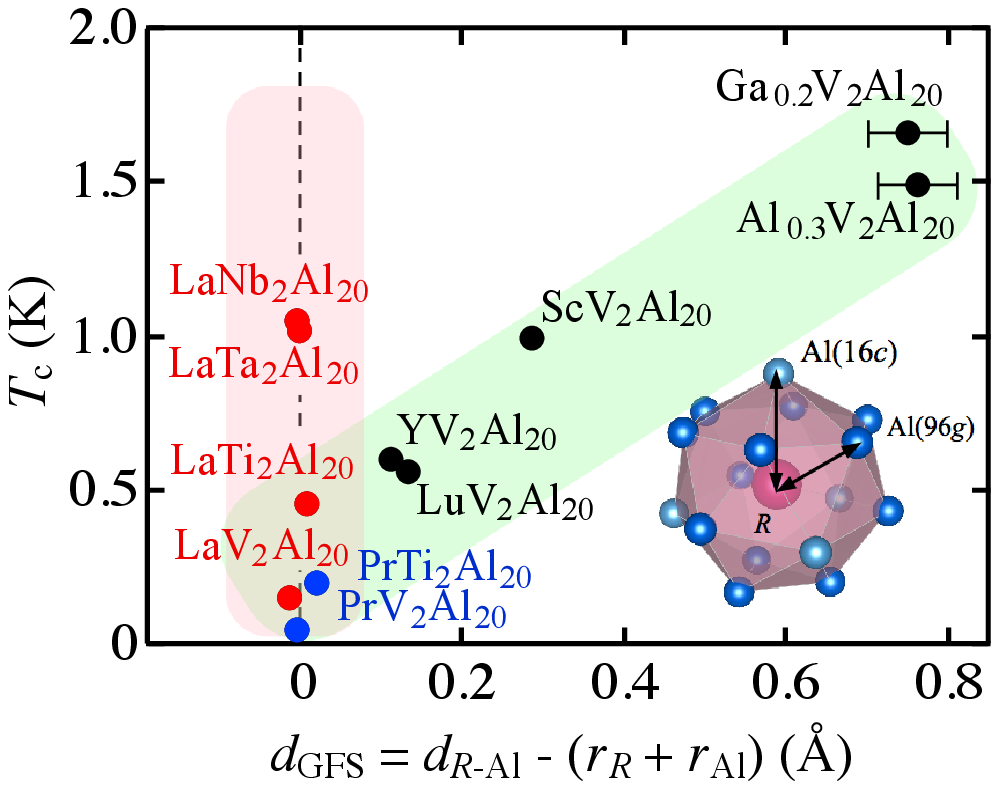}
\end{center}
\caption{(Color online) $\Tc$ vs $d_{\rm GFS} \equiv \it d_{R \rm \--Al}-(\it r_{R}+\it r_{\rm Al})$ quantifying the ``guest free space" of nonmagnetic cage-center $R$ ions for $\it RTr_{2}$Al$_{20}$ (see text for details). %
Inset shows the $R(8a)$-Al$(96g, 16c)$ cage.
We adopt $r_{\rm Al}$ = 1.18 $\rm\AA$ from the proposed 1.21(4) $\rm\AA$~\cite{Cordero_DT_08}.
This figure demonstrates that nonmagnetic $\it RTr_{2}$Al$_{20}$ superconductors are classified into two groups, i.e., (A) $d_{\rm GFS} \ne 0$ and $\Tc$ correlates with $d_{\rm GFS}$, and (B) $d_{\rm GFS} \simeq 0$ and $\Tc$ seems to be governed by other factors.
Note that superconductors PrTi$_2$Al$_{20}$~\cite{Sakai_JPSJ_12_PrTi2Al20} and PrV$_2$Al$_{20}$~\cite{Tsujimoto_PRL_14}, and field-insensitive HF compounds Sm$Tr_2$Al$_{20}$ ($Tr=$ Ti, V, Cr, and Ta)~\cite{Higashinaka_JPSJ_11_SmTi2Al20, Sakai_PRB_11, Yamada_JPSJ_13} also have $d_{\rm GFS} \simeq 0$. %
}
\label{figTc}
\end{figure}

Using the $\chi'$ and $C$ data measured in applied fields, the SC phase diagram is constructed as shown in Fig.~\ref{figHTdiagram}.
The two sets of the SC boundary data agree well with each other.
$H_{\rm c2}(0)$ is much lower than the Pauli-limiting field $H_{\rm P}$ = (1.84$\times 10^4$ Oe/K) $T_{\rm c}$~\cite{Clogston}, suggesting that $H_{\rm c2}(0)$ is determined by the orbital depairing effect. 
The temperature dependence of $H_{\rm c2}$ can be well described by the Werthamer-Helfand-Hohenberg (WHH) clean-limit expression~\cite{HW_PR_66, WHH_PR_66}, as shown by the solid curves in Fig.~\ref{figHTdiagram}.
In this model, $H_{\rm c2}$ at $T=0$ satisfies
\begin{equation}
H_{\rm c2}(0) = -0.73\times\frac{dH_{\rm c2}}{dT}|_{T=T_{\rm c}}T_{\rm c} = \frac{\phi_{0}}{2 \pi \xi_{\rm GL}^2},
\label{eqWHH}
\end{equation}
where $\phi_{0}$ and $\xi_{\rm GL}$ are the quantum magnetic flux and the Ginzburg-Landau (GL) coherence length, respectively.
The GL parameter $\kappa_{\rm GL}$, which is equal to the Maki parameter~\cite{Maki} $\kappa_2(T \to T_{\rm c})$, is determined using the thermodynamic relation~\cite{Serin}: %
\begin{equation}
\frac{\Delta C_{\rm vol}}{T}\vert_{T=T_{\rm c}} = (\frac{dH_{\rm c2}}{dT}|_{T=T_{\rm c}})^2 \frac{1}{4 \pi (2 \kappa_2^2-1) \beta_{\rm A}},
\label{kappa2}
\end{equation}
where $\Delta C_{\rm vol}$ is measured per unit volume [unit: erg/(K cm$^3$)], and $\beta_{\rm A}=1.16$ for a triangular vortex lattice.
The thermodynamic critical field $H_{\rm c}(0) = \alpha \sqrt{(6/\pi) \gamma_{\rm vol}} T_{\rm c}$~\cite{Johnson_SST_13}, the London penetration depth $\lambda_{\rm L} = \kappa_{\rm GL} \xi_{\rm GL}$, and the lower critical field $H_{\rm c1} = H_{\rm c}(0) \ln \kappa_{\rm GL}/(\sqrt{2} \kappa_{\rm GL})$ are also calculated. 
The obtained characteristic parameters are summarized in Table~\ref{tableSCpara}.

The SC properties of LaTi$_{2}$Al$_{20}$, which are significantly different from the others, strongly suggest that it is a type-I superconductor.
(1) The SC phase boundary is well described by $H_{\rm c}(T)=H_{\rm c0}[1-(T/T_{\rm c})^s]$ rather than the WHH model; the best fit parameters are $s=2.6$ ($s=2$ is expected for a conventional type-I SC) and $H_{\rm c0}=27$ Oe, which agrees well with $H_{\rm c}(0)=25$ Oe.
(2) The Meissner state ($\chi'=$ const.) dominates and extends over a large $H$ range in the SC phase (see Fig.~\ref{figchiac}).
(3) Positive peaks appear in $\chi'$ at the SC phase boundary (i.e., $\partial M/\partial H>0$), which are ascribable to the "differential paramagnetic effect (DPE)" often observed in type-I superconductors~\cite{HeinFalge, Yonezawa, smallDPE}.
Note that most known type-I superconductors are pure metals and only a handful are reported among compounds and alloys.

The electron-phonon coupling constant $\lambda_{\rm e\textendash ph}$ is obtained using the modified McMillan's formula
\begin{equation}
\lambda_{\rm e\textendash ph} = \frac{1.04 + \mu^{*}\ln(\frac{\it\Theta_{\rm D}}{1.45T_{\rm c}})}{(1-0.62\mu^{*})\ln(\frac{\it\Theta_{\rm D}}{1.45T_{\rm c}}) - 1.04},
\label{eqmodMcMillan}
\end{equation}
where the Coulomb coupling constant $\mu^{*}$ is assumed to be 0.13~\cite{McMillan_PR_68} and the Debye temperature $\it\Theta_{\rm D}$ is obtained from $\it\Theta_{\rm D} = \sqrt[3]{\rm(12/5)\pi^{4}\it nR/\beta}$ ($n=23$ is the number of atoms per formula unit and $R$ is the gas constant).
The fact that $\lambda_{\rm e\textendash ph}$ ranges between 0.34 and 0.42 is consistent with the above-mentioned weak-coupling nature of the superconductivity in $\it RTr_{2}$Al$_{20}$.

One of the important features here is that the large distribution of $\Tc$( = 0.15-1.05 K) is mainly caused by the different $\lambda_{\rm e\--ph}$( = 0.34-0.42) in La$\it Tr\rm_{2}$Al$_{20}$.
Note that $\Tc$ correlates strongly with $\lambda_{\rm e\--ph}$ but not with $\it\Theta_{\rm D}$, which is determined below 5 K.
This fact suggests that acoustic phonons are not playing a major role in the Cooper pairing.

For LaTi$_{2}$Al$_{20}$, the mass enhancement factor $\gamma/\gamma_{\rm cal}$ is 1.26, which is slightly smaller than the expected $(1+\lambda_{\rm e\textendash ph}) = 1.38$.
These values are in good agreement with a de Haas-van Alphen (dHvA) experiment, which showed that the cyclotron mass is approximately 1.2 times larger than that obtained from a band structural calculation~\cite{Nagashima_JPSCP_14, mass}.

Here, we examine the crystallographic features of the Al$_{16}$ cage, which includes a La ion at the center ($8a$ site).
For $\it RTr\rm_{2}$Al$_{20}$ with $\it R\rm=$ lanthanide ions, it is considered that a guest $\it R$ ion does not have a free space around it inside the Al$_{16}$ cage as suggested by the fact that the lattice parameter $a$ depends almost linearly on the covalent radius of the lanthanide ion $r_{R}$~\cite{Winiarski_PRB_16}.
On the contrary, for $R=$ Ga, Al, Sc and Lu, $a$ (= 14.49-14.51 $\rm\AA$) is almost independent of $r_{R}$, which is smaller than those for lanthanide ions, suggesting the existence of a free space around the $R$ ions.
Some of the anomalous SC properties observed for $R=$ Ga, Al, Sc and Lu are attributable to the ``rattling" anharmonic low-frequency vibrations of the $R$ ions, which are considered to couple strongly with conduction electrons and to enhance $\Tc$~\cite{Hiroi_JPSJ_12, Onosaka_JPSJ_12, Safarik_PRB_12, Koza_PCCP_14, Winiarski_PRB_16}.

We introduce a parameter to quantify the ``guest free space" as $d_{\rm GFS} \equiv d_{R \rm \--Al}-(\it r_{R}+\it r_{\rm Al})$, where $d_{R \rm \--Al} \equiv (12 d_{R \rm \--Al(96g)}+4 d_{R \rm \--Al(16c)})/16$ is the average distance between $R$ and Al in a cage (see the inset of Fig.~\ref{figTc}), and $r_{R}$ and $r_{\rm Al}$ are the covalent radii for $\it R$ and Al ions, respectively~\cite{Cordero_DT_08}.
$d_{R \rm \--Al}$ is calculated using the results of the single-crystal X-ray diffraction analysis shown in Table~\ref{tablestruct}.
$\Tc$ is plotted as a function of $\it d_{\rm GFS}$ in Fig.~\ref{figTc} for $\it RTr_{2}$Al$_{20}$ with nonmagnetic $\it R$ ions.
For $\it Tr$ = V, $\Tc$ increases almost linearly with $\it d_{\rm GFS}$.
This behavior agrees well with the interpretation that $\Tc$ is enhanced by the ``rattling" anharmonic vibration modes of Ga, Al, Sc, and Lu ions.
In contrast, the four data points of the present study fall almost into a vertical line with $\it d_{\rm GFS} \rm\simeq 0$, indicating that these La compounds do not have guest free space and the large $\Tc$ distribution is not associated with the La ion oscillations.
We speculate that the $\Tc$ distribution may be caused by differences in (i) anharmonic large-amplitude vibration modes of Al(16c) ions and/or (ii) characteristics of the $d$ electrons of $\it Tr$ ions ($e.g.$, stronger spin-orbit interaction for $4d$ and $5d$ than $3d$).

In summary, we have grown single crystals of {\LaTAl} ($\it Tr\rm$ = Ti, V, Nb, and Ta) and have revealed that they are new weak-coupling superconductors.
The single-crystal X-ray diffraction analyses have clarified that they form a group of superconductors characterized by no ``guest free space" for the cage-center La ions.

\acknowledgment
This work was supported by MEXT/JSPS KAKENHI grants 15H03693, 15H05884, 15J07600, and 15K05178.




\begin{thebibliography}{9}
\bibitem{Torikachvili_PNAS_07}M. S. Torikachvili, S. Jia, E. D. Mun, S. T. Hannahs, R. C. Black, W. K. Neils, D. Martien, S. L. Bud$^{\prime}$ko, and P. C. Canfield, Proc. Natl. Acad. Sci. U.S.A. \bf104\rm, 9960 (2007).
\bibitem{Higashinaka_JPSJ_11_SmTi2Al20}R. Higashinaka, T. Maruyama, A. Nakama, R. Miyazaki, Y. Aoki, and H. Sato, J. Phys. Soc. Jpn. \bf80\rm, 093703 (2011).
\bibitem{Sakai_PRB_11}A. Sakai and S. Nakatsuji, Phys. Rev. B \bf84\rm, 201106(R) (2011).
\bibitem{Yamada_JPSJ_13}A. Yamada, R. Higashinaka, R. Miyazaki, K. Fushiya, T. D. Matsuda, Y. Aoki, W. Fujita, H. Harima, and H. Sato, J. Phys. Soc. Jpn. \bf82\rm, 123710 (2013).
\bibitem{Sakai_JPSJ_11}A. Sakai and S. Nakatsuji, J. Phys. Soc. Jpn. \bf80\rm, 063701 (2011). %
\bibitem{Onimaru_JPSJ_16}T. Onimaru and H. Kusunose, J. Phys. Soc. Jpn. \bf85\rm, 082002 (2016).
\bibitem{Yoshida_JPSJ_17}T. Yoshida, Y. Machida, K. Izawa, Y. Shimada, N. Nagasawa, T.Onimaru, T. Takabatake, A. Gourgout, A. Pourret, G. Knebel, and J.-P.Brison, J. Phys. Soc. Jpn. \bf86\rm, 044711 (2017).
\bibitem{Higashinaka_JPSJ_17}R. Higashinaka, A. Nakama, R. Miyazaki, J. Yamaura, H. Sato, and Y. Aoki, J. Phys. Soc. Jpn. \bf86\rm, 103703 (2017).
\bibitem{Onimaru_JPSJ_11} T. Onimaru, K. T. Matsumoto, Y. F. Inoue, K. Umeo, T. Sakakibara, Y. Karaki, M. Kubota, and T. Takabatake, Phys. Rev. Lett. \bf106\rm, 177001 (2011). 
\bibitem{Sakai_JPSJ_12_PrTi2Al20}A. Sakai, K. Kuga, and S. Nakatsuji, J. Phys. Soc. Jpn. \bf81\rm, 083702 (2012).%
\bibitem{Matsubayashi_PRL_12}K. Matsubayashi, T. Tanaka, A. Sakai, S. Nakatsuji, Y. Kubo, and Y. Uwatoko, Phys. Rev. Lett. \bf109\rm, 187004 (2012). %
\bibitem{Tsujimoto_PRL_14}M. Tsujimoto, Y. Matsumoto, T. Tomita, A. Sakai, and S. Nakatsuji, Phys. Rev. Lett. \bf113\rm, 267001 (2014).%
\bibitem{Hiroi_JPSJ_12} Z. Hiroi, A. Onosaka, Y. Okamoto, J. Yamaura, and H. Harima, J. Phys. Soc. Jpn. \bf81\rm, 124707 (2012).
\bibitem{Safarik_PRB_12}D. J. Safarik, T. Klimczuk, A. Llobet, D. D. Byler, J. C. Lashley, J. R. O'Brien, and N. R. Dilley, Phys. Rev. B \bf85\rm, 014103 (2012).
\bibitem{Onosaka_JPSJ_12}A. Onosaka, Y. Okamoto, J. Yamaura, and Z. Hiroi, J. Phys. Soc. Jpn. \bf81\rm, 023703 (2012).
\bibitem{Koza_PCCP_14}M. M. Koza, A. Leithe-Jasper, E. Sischka, W. Schnelle, H. Borrmann, H. Mutka, and Y. Grin, Phys. Chem. Chem. Phys. \bf16\rm, 27119 (2014).
\bibitem{Winiarski_PRB_16} M. J. Winiarski, B. Wiendlocha, M. Sternik, P. Wi$\acute{\rm s}$niewski, J. R. O'Brien, D. Kaczorowski, and T. Klimczuk, Phys. Rev. B \bf93\rm, 134507 (2016).
\bibitem{Onimaru_JPSJ_10}T. Onimaru, K. T. Matsumoto, Y. F. Inoue, K. Umeo, Y. Saiga, Y. Matsushita, R. Tamura, K. Nishimoto, I. Ishii, T. Suzuki, and T. Takabatake, J. Phys. Soc. Jpn. \bf79\rm, 033704 (2010).
\bibitem{Wakiya_JPSJ_17} K. Wakiya, T. Onimaru, K. Matsumoto, Y. Yamane, N. Nagasawa, K. Umeo, S. Kittaka, T. Sakakibara, Y. Matsushita, and T. Takabatake, J. Phys. Soc. Jpn. \bf86\rm, 034707 (2017).
\bibitem{Hasegawa_JPCS_12}T. Hasegawa, N. Ogita, and M. Udagawa: J. Phys. Conf. Ser. \bf391 \rm (2012) 012016.
\bibitem{Wakiya_PRB_16}K. Wakiya, T. Onimaru, S. Tsutsui, T. Hasegawa, K. T. Matsumoto, N.
Nagasawa, A. Q. R. Baron, N. Ogita, M. Udagawa, and T. Takabatake, Phys. Rev. \bf93\rm, 064105 (2016).
\bibitem{chi_ac} The susceptometer consists of a magnetizing (primary) coil and two identical pick-up (secondary) coils connected in series opposition. The generated ac voltage across the pick-up coils, which is proportional to $\chi_{\rm ac}$ of a sample centered in one of the secondary coils, was measured using an ac resistance bridge LR-700 (Linear Research Inc.). The absolute values of $\chi_{\rm ac}$ is calculated considering the geometry of the coils [see, \textit{e.g.}, R. B. Goldfarb and J. V. Minervini, Rev. Sci.Instrum. \bf55\rm, 761 (1984)].
\bibitem{Sheldrick_SHELX-97_97}G. M. Sheldrick, Acta Cryst. \bf A64 \rm, 112 (2008).

\bibitem{Niemann_JSSC_95}S. Niemann and W. Jeitschko, J. Solid State Chem. \bf114\rm, 337 (1995).
\bibitem{Thiede_JAAC_98}V. M. T. Thiede, W. Jeitschko, S. Niemann, and T. Ebel, J. Alloys Compd. \bf267\rm, 23 (1998).

\bibitem{Nasch_ZNB_97}T. Nasch, W. Jeitschko, and U. C. Rodewald, Z. Naturforsch. B \bf52\rm, 1023 (1997).
\bibitem{Kangas_JSSC_12}M. J. Kangas, D. C. Schmitt, A. Sakai, S. Nakatsuji, J. Y. Chan, J. Solid State Chem. \bf196\rm, 274 (2012).

\bibitem{Padamsee_JLTP_73}H. Padamsee, J. E Neighbor, and C. A. Shiffman, J. Low Temp. Phys. \bf12\rm, 387 (1973).
\bibitem{Johnson_SST_13}D. C. Johnston, Supercond. Sci. Technol. \bf26\rm, 115011 (2013).

\bibitem{HW_PR_66} E. Helfand and N. R. Werthamer, Phys. Rev. \bf147\rm, 288 (1966).
\bibitem{WHH_PR_66} N. R. Werthamer, E. Helfand, and P. C. Hohenberg, Phys. Rev. \bf147\rm, 295 (1966).
\bibitem{SJ_dG_PL_64}D. Saint-James and P. G. Gennes, Phys. Lett. \bf7\rm, 306 (1963).

\bibitem{Clogston} A. M. Clogston, Phys. Rev. Lett. \bf9\rm, 266 (1962).

\bibitem{Maki} K. Maki, Physics (Long Island City, N.Y.) \bf1\rm, 21 (1964). %

\bibitem{Serin} B. Serin: in {\it Superconductivity}, ed. R. D. Parks (Marcel Dekker, New York, 1969) Vol. 2, Chap. 15. %


\bibitem{HeinFalge} R. A. Hein and R. L. Falge, Jr., Phys. Rev. \bf123\rm, 407 (1961). %
\bibitem{Yonezawa} S. Yonezawa and Y. Maeno, Phys. Rev. B \bf72\rm, 180504(R) (2005). %
\bibitem{smallDPE} The rather small size of the DPE peak in LaTi$_{2}$Al$_{20}$ may possibly be due to distribution of the diamagnetic field in the crystals and/or the existence of domain wall pinning in the intermediate state. %

\bibitem{McMillan_PR_68} W. L. McMillan, Phys. Rev. \bf167\rm, 331 (1968).


\bibitem{Nagashima_JPSCP_14} S. Nagashima, T. Nishiwaki, A. Otani, M. Sakoda, E. Matsuoka, H. Harima, and H. Sugawara, JPS Conf. Proc. \bf3\rm, 011019 (2014).


\bibitem{mass} As shown in Table~\ref{tableSCpara}, $dH_{\rm c2}/dT \vert_{T=T_{\rm c}}$ has a large distribution in nonmagnetic $R \it Tr\rm_{2}$Al$_{20}$.
This quantity is expected to be proportional to the mass enhancement factor $(\gamma/\gamma_{\rm cal})^2$.
To check this relation, theoretical calculations for $\gamma_{\rm cal}$ are desired. %


\bibitem{Cordero_DT_08} B. Cordero, V. G$\acute{\rm o}$mez, A. E. Platero-Prats, M. Rev$\acute{\rm e}$s, J. Echeverr$\acute{\imath}$a, E. Cremades, F. Barrag$\acute{\rm a}$n, and S. Alvarez, Dalton Trans. \bf21\rm, 2832 (2008).

\bibitem{Tanaka_JPSJ_11} T. Tanaka and Y. Kubo, J. Phys. Soc. Jpn. \bf80\rm, SA125 (2011).
\bibitem{Sakai_JPSJ_12}A. Sakai and S. Nakatsuji, J. Phys. Soc. Jpn. \bf81\rm, SB049 (2012).


\end{thebibliography}
\end{document}